\documentclass[12pt]{article}
\usepackage{amsmath}
\usepackage{graphicx}
\usepackage{amsfonts}
\usepackage{amssymb}
\newtheorem{theorem}{Theorem}[section]

\newtheorem{lemma}[theorem]{Lemma}

\newtheorem{remark}[theorem]{Remark}
\newenvironment{proof}[1][Proof]{\textsc{#1.} }{\ \rule{0.5em}{0.5em}}
\numberwithin{equation}{section}

\begin{document}

\title{Bianchi - Euler system for relativistic fluids\\and Bel - Robinson type energy.}
\author{Yvonne Choquet-Bruhat and James W. York}
\maketitle
\begin{abstract}
We write a first order symmetric hyperbolic system coupling the Riemann tensor
with the dynamical acceleration of a prefect relativistic fluid. We determine
the associated, coupled, Bel - Robinson type energy, and the integral equality
that it satisfies.
\end{abstract}

\textbf{Syst\`{e}me de Bianchi-Euler pour un fluide relativiste, et
\'{e}nergie de type Bel-Robinson.}

\textbf{R\'{e}sum\'{e}.} On \'{e}crit un syst\`{e}me sym\'{e}trique
hyperbolique satisfait par le tenseur de Riemann de l'espace temps et
l'acc\'{e}l\'{e}ration dynamique d'un fluide parfait relativiste. On
d\'{e}termine l'\'{e}nergie du type Bel-Robinson correspondante, et
l'\'{e}galit\'{e} int\'{e}grale qu'elle satisfait.

\section{Introduction.}

The effective strength of the gravitational field lies in the Riemann tensor
of the spacetime metric. Its evolution is governed by the so - called higher
order equations (Bel 1958, Lichnerowicz 1964), deduced from the Bianchi
identities. The system satisfied by the trace free part of the Riemann tensor,
the Weyl tensor, was some time ago recognized as a linear, first order
symmetric hyperbolic system (FOSH), with constraints, homogeneous in vacuum.
See H. Friedrich (1996) and references therein. The evolution equations for
the Riemann tensor itself have also been written (CB-Yo 1997) as a
FOSH\ system, made explicit in terms of four two- tensors, introduced by Bel
(1958) in a general setting, the electric and magnetic gravitational fields
and corresponding inductions relative to a Cauchy adapted frame. In the
presence of sources the higher order equations, now renamed \textit{Bianchi
equations}, are no longer homogeneous; their right hand sides are linear in
the covariant derivative of the stress energy tensor of the sources.

In this article we consider the case of perfect fluid sources. The Einstein
equations with such sources have long ago (CB 1958) been proved to be a well
posed Leray - hyperbolic system (with constraints). The fluid equations have
also been written as a FOSH system (in special relativity, K.O. Friedrichs
1969; in general relativity Ruggeri and Strumia 1981, Anile 1982, and Rendall
1992 who proved a theorem valid also for isolated fluid bodies with special
equations of state). It seemed however interesting to have a system of
equations which would be a FOSH system both for the gravitational field,
namely the Riemann tensor of space time, and the fluid variables. Such a
system has been written in lagrangian variables (that is, in a frame whose
timelike axis is tangent to the fluid flow lines) by H. Friedrich 1998, who
used the Weyl tensor and by CB-Yo 2001, using directly the Riemann tensor. In
this paper we use eulerian variables, that is a frame adapted to the usual 3+1
slicing of the space time, with time axis orthogonal to the space slices,
which we call a Cauchy adapted frame. We obtain a FOSH system for the Riemann
tensor and the dynamical fluid acceleration. The energy corresponding to the
FOSH\ system that we obtain, and call the Bianchi - Euler system, is the sum
of the usual Bel Robinson energy of the gravitational field, and the dynamical
acceleration energy of the fluid. It is not conserved in general - no more
than the Bel - Robinson energy in vacuum - but its evolution can be controlled
through the equations we obtain.

\section{Einstein equations with fluid sources.}

The Einstein equations with source a stress energy tensor are:
\[
R_{\alpha\beta}=\rho_{\alpha\beta},\text{ \ \ }\rho_{\alpha\beta}\equiv
T_{\alpha\beta}-\frac{1}{2}g_{\alpha\beta}T
\]

The stress energy tensor of a perfect relativistic fluid is
\[
T_{\alpha\beta}\equiv(\mu+p)u_{\alpha}u_{\beta}+pg_{\alpha\beta}%
\]
with $u^{\alpha}$ the unit kinematical velocity satisfying $u^{\alpha
}u_{\alpha}=-1;$ and $\mu,p,S$ the specific energy, pressure and entropy.
These thermodynamic quantities are assumed to be positive. They are linked by
an equation of state:
\begin{equation}
\mu=\mu(p,S)
\end{equation}
The enthalpy index $f$ of the fluid is given by the identity
\begin{equation}
f(p,S)\equiv\chi\exp\int_{p_{0}}^{p}\frac{dp}{\mu(p,S)+p},
\end{equation}
where $\chi$ is a constant such that $f^{2}$ has the dimensions of an energy
density\footnote{{\footnotesize That is, in general relativity where time and
length have the same dimension and mass-energy the dimension of a length, f
and therefore }$\chi$ {\footnotesize have dimension of the inverse of a
length.}}.

The dynamical velocity $C^{\alpha}$, a vector tangent to the flow lines,
incorporates information on the kinematic velocity $u^{\alpha}$ and the
thermodynamic quantities. It is defined by
\begin{equation}
C^{\alpha}\equiv fu^{\alpha},
\end{equation}
hence it satisfies the relation
\begin{equation}
C^{\alpha}C_{\alpha}=-f^{2}.
\end{equation}
In terms of the dynamical velocity the equations of a perfect fluid in an
arbitrary frame are
\begin{equation}
C^{\alpha}\nabla_{\alpha}C_{\beta}+f\partial_{\alpha}f\equiv C^{\alpha}%
(\nabla_{\alpha}C_{\beta}-\nabla_{\beta}C_{\alpha})=0
\end{equation}
and, with $\mu_{p}^{\prime}(p,S)\equiv\partial\mu(p,S)/\partial p,$ a given
function of $p$ and $S,$%
\begin{equation}
\nabla_{\alpha}C^{\alpha}+(\mu_{p}^{\prime}-1)\frac{C^{\alpha}C^{\beta}%
}{C^{\lambda}C_{\lambda}}\nabla_{\alpha}C_{\beta}=0
\end{equation}%
\begin{equation}
C^{\alpha}\nabla_{\alpha}S=0.
\end{equation}

In these equations the unknowns are the four components of the vector
$C^{\alpha},$ and the scalar $S.$ The specific pressure $p$ is a known
function of $f$ (i.e. of $C^{\alpha},$ by 2.4) and of $S$%
\begin{equation}
p\equiv p(C^{\alpha}C_{\alpha},S)
\end{equation}
determined by inverting the relation 2.2.

There seems to be more equations than unknowns, but the equations 2.5. are not
independent, because they satisfy the following identity:
\[
C^{\alpha}C^{\beta}(\nabla_{\alpha}C_{\beta}-\nabla_{\beta}C_{\alpha}%
)\equiv0.
\]
The equation 2.7 says that $S$ is constant along the flow lines, hence
constant in spacetime if constant initially. We suppose, to simplify what
follows below, that $S$ is constant. Removing this hypothesis introduces no
essential difficulty, but one has to add as new unknowns the derivatives
$\partial_{i}S,$ and the equations obtained from 2.7 by taking a derivative
$\nabla_{i}.$ Our results are valid as they stand for a barotropic fluid,
because then $\mu\equiv\mu(p).$

\section{Bianchi equations.}

We now work in a Cauchy adapted frame, that is, a frame with its timelike axis
orthogonal to the space slices. The spacetime metric takes then the usual 3+1
form
\[
ds^{2}=-N^{2}dt^{2}+g_{ij}(dx^{i}+\beta^{i}dt)(dx^{j}+\beta^{j}dt).
\]
The derivatives $\partial_{\alpha}$ are the Pfaff derivatives in the coframe
$\theta^{0}=dt,$ \ \ $\theta^{i}=dx^{i}+\beta^{i}dt,$ that is
\[
\partial_{0}=\frac{\partial}{\partial t}-\beta^{i}\frac{\partial}{\partial
x^{i}},\text{ \ \ }\partial_{i}=\frac{\partial}{\partial x^{i}}.
\]

We have written in CB.Yo 1997 the Bianchi equations satisfied by the Riemann
tensor as a FOSH system
\begin{equation}
\nabla_{0}R_{hk,0j}+\nabla_{k}R_{0h,0j}-\nabla_{h}R_{0k,0j}=0
\end{equation}
and
\begin{equation}
\nabla_{0}R_{:::i,0j}^{0}+\nabla_{h}R_{:::i,0j}^{h}=J_{i,0j}\equiv\nabla
_{0}\rho_{ji}-\nabla_{j}\rho_{0i}%
\end{equation}
The equations 3.1. and 3.2. are for each given pair ($0j)$ a first order
symmetric system, hyperbolic relative to the space sections for the components
$R_{hk,0j\text{ }}$ and $R_{0h,0j}$ because the matrix $M^{0}$ of the
coefficients of the derivatives $\partial_{0}$ is the unit matrix, and the
matrix $M^{t}$ of the coefficients of the derivatives $\frac{\partial
}{\partial t}$ is indentical to $M^{0}.$

Analogous results hold for the components $R_{hk,ij\text{ }}$ and $R_{0h,ij}. $

To the Bianchi system is associated its Bel - Robinson energy density on a
space slice, namely
\[
\mathcal{B}\equiv{\frac{1}{2}}(|\mathbf{E}|^{2}+|\mathbf{H}|^{2}%
+|\mathbf{D}|^{2}+|\mathbf{B}|^{2})
\]
where $\mathbf{E,H,D,B}$ are the electric and magnetic fields and inductions
space 2- tensors associated with the Riemann tensor\footnote{$\eta_{ijk}$
{\footnotesize the volume form,
%TCIMACRO{\TEXTsymbol{\vert}}%
%BeginExpansion
$\vert$%
%EndExpansion%
%TCIMACRO{\TEXTsymbol{\vert}}%
%BeginExpansion
$\vert$%
%EndExpansion
and }$\bar{\nabla},$ {\footnotesize the norm and} {\footnotesize covariant
derivative, are defined by the space metric g}$_{ij}.${\footnotesize .}}:
\[
E_{ij}\equiv R^{0}{}_{i,0j},\text{ \ }D_{ij}\equiv\frac{1}{4}\eta_{ihk}%
\eta_{jlm}R^{hk,lm},\text{ }.
\]%
\[
\text{\ }H_{ij}\equiv\frac{1}{2}N^{-1}\eta_{ihk}R^{hk},_{oj},\text{ \ }%
B_{ji}\equiv\frac{1}{2}N^{-1}\eta_{ihk}R_{0j},^{hk}.
\]
This energy satisfies the equality
\begin{equation}
{\frac{1}{2}\partial}{_{0}}(|\mathbf{E}|^{2}+|\mathbf{H}|^{2}+|\mathbf{D}%
|^{2}+|\mathbf{B}|^{2})+\bar{\nabla}_{h}\{N\eta^{lh}{}_{i}(E^{ij}H_{lj}%
-B^{ij}D_{lj}\}=Q(\mathbf{E},\mathbf{H},\mathbf{D},\mathbf{B})+\mathcal{S}%
\nonumber
\end{equation}
where $Q$ is a quadratic form in $\mathbf{E,...B}$ with coefficients
${\mathbf{\bar{\nabla}}}N$ and $N\mathbf{K,}$ $\mathbf{K}$ extrinsic curvature
of the spaceslices. The source term $\mathcal{S}$, zero in vacuum, is
\begin{equation}
\mathcal{S}\equiv J_{0ij}E^{ij}-{\frac{1}{2}}NJ_{lmi}\eta_{h}\mathstrut
^{lm}B^{ih}.
\end{equation}

The tensor $\rho_{\alpha\beta}$ for a perfect fluid is given in terms of
$C_{\alpha}$ by
\begin{equation}
\rho_{\alpha\beta}\equiv(\mu+p)f^{-2}C_{\alpha}C_{\beta}+\frac{1}{2}%
g_{\alpha\beta}(\mu-p)
\end{equation}
The sources of the Bianchi equations are therefore of the first order in
$C_{\alpha},$ linear in the derivatives $\nabla_{\alpha}C_{\beta}.$

\section{Equations for $\nabla C.$}

The dynamical acceleration $\nabla C$ satisfies the following equations
obtained by covariant differentiation of 2.5. and 2.6, and use of the Ricci
identities: \ \ \ \
\begin{equation}
M_{\gamma\beta}\equiv C^{\alpha}(\nabla_{\alpha}C_{\gamma\beta}-\nabla_{\beta
}C_{\gamma\alpha})+a_{\gamma\beta}=0
\end{equation}
and
\begin{equation}
g^{\alpha\beta}\nabla_{\alpha}C_{\gamma\beta}+(\mu_{p}^{\prime}-1)\frac
{C^{\alpha}C^{\beta}}{C^{\lambda}C_{\lambda}}\nabla_{\alpha}C_{\gamma\beta
}+b_{\gamma}=0
\end{equation}
where we have set
\begin{equation}
C_{\gamma\beta}\equiv\nabla_{\gamma}C_{\beta},
\end{equation}%
\begin{equation}
a_{\gamma\beta}\equiv C_{\gamma}{}^{\alpha}(C_{\alpha\beta}-C_{\beta\alpha
})+C^{\alpha}C_{\lambda}R_{\gamma\alpha,\beta}^{.........\lambda}%
\end{equation}%
\begin{equation}
b_{\gamma}\equiv-R_{\gamma\lambda}C^{\lambda}+\nabla_{\gamma}\{(\mu
_{p}^{\prime}-1)\frac{C^{\alpha}C^{\beta}}{C^{\lambda}C_{\lambda}}%
\}C_{\alpha\beta}%
\end{equation}
The last term in $b_{\gamma}$ is a quadratic form in $C_{\alpha\beta}$ whose
coefficients are functions of the $C^{\alpha}$ and $S.$ These functions can be
computed by using the identity (we now use $S=$ constant)
\[
\nabla_{\gamma}\mu_{p}^{\prime}\equiv\mu_{p^{2}}^{\prime\prime}\partial
_{\gamma}p,\text{ \ \ }\mu_{p^{2}}^{\prime\prime}(p,S)\equiv\frac{\partial
\mu^{\prime}(p,S)}{\partial p}.
\]
with by the definition of $f$ and the identity 2.4 it holds that
\[
\partial_{\gamma}p=(\mu+p)f^{-1}\partial_{\gamma}f=-(\mu+p)(C^{\lambda
}C_{\lambda})^{-1}C^{\alpha}C_{\gamma\alpha}.
\]
We complete the computation by using 2.1 and 2.8.

The equations 4.1. are not independent, because they satisfy the identities
\begin{equation}
C^{\beta}M_{\gamma\beta}\equiv0.
\end{equation}
The equations 4.1 and 4.2. are not a well posed system. Instead of the
$4\times4$ equations 4.1 we consider\footnote{{\footnotesize An analogous
procedure is used for the symmetrization of the Euler equations in K.O.
Friedrichs 1969\ and in Rendall 1992.}} the $4\times3$ ones:
\begin{equation}
\tilde{M}_{\gamma i}\equiv M_{\gamma i}-\frac{C_{i}}{C_{0}}M_{\gamma0}=0
\end{equation}
The terms in derivatives of $C_{\gamma\lambda}$ in these equations can be
written in the following form:
\begin{equation}
C^{\alpha}\partial_{\alpha}(C_{\gamma i}-\frac{C_{i}}{C_{0}}C_{\gamma
0})-(\partial_{i}-\frac{C_{i}}{C_{0}}\partial_{0})(C^{\alpha}C_{\gamma\alpha})
\end{equation}

\begin{lemma}
The system 4.2,4.7. is equivalent to a FOS (First Order Symmetric) system
\ for $C_{\gamma\alpha}$ with coefficients being functions of the Riemann
tensor, the connection and the dynamical velocity $C_{\lambda},$ but not of
\ their derivatives.
\end{lemma}

\begin{proof}
The system is quasi diagonal by blocks, each block corresponding to a given
value of the index $\gamma.$ We will write the principal operator of a block
by omitting this index. We set
\begin{equation}
U_{i}\equiv C_{\gamma i}-\frac{C_{i}}{C_{0}}C_{\gamma0},\text{ \ \ \ }%
U_{0}\equiv C^{\alpha}C_{\gamma\alpha}%
\end{equation}
and we define the differential operators $\tilde{\partial}_{\alpha}$ as
follows:
\begin{equation}
\tilde{\partial}_{0}\equiv C^{\alpha}\partial_{\alpha},\text{ \ \ \ }%
\tilde{\partial}_{i}=\partial_{i}-\frac{C_{i}}{C_{0}}\partial_{0}%
\end{equation}
The principal terms (derivatives of $C_{\gamma\alpha})$ in the equations 4.7
with index $\gamma$ are
\begin{equation}
\tilde{\partial}_{0}U_{i}-\tilde{\partial}_{i}U_{0}.
\end{equation}
We have by inverting 4.9:
\[
C_{\gamma0}\equiv\frac{C_{0}(U_{0}-C^{i}U_{i})}{C^{\lambda}C_{\lambda}}%
\]%
\[
C_{\gamma i}\equiv U_{i}+\frac{C_{i}(U_{0}-C^{j}U_{j})}{C^{\lambda}C_{\lambda
}}%
\]
The principal terms of 4.2. read, using the above formulae
\begin{equation}
\frac{\mu_{p}^{\prime}C^{\alpha}\partial_{\alpha}U_{0}}{C^{\lambda}C_{\lambda
}}+(g^{ij}-\frac{C^{i}C^{j}}{C^{\lambda}C_{\lambda}})\partial_{i}U_{j}%
-\frac{C^{0}C^{i}}{C^{\lambda}C_{\lambda}}\partial_{0}U_{i}%
\end{equation}
\end{proof}

We introduce the positive definite (if $C$ is timelike) quadratic form
\begin{equation}
\tilde{g}^{ij}\equiv g^{ij}-\frac{C^{i}C^{j}}{C^{\lambda}C_{\lambda}}.
\end{equation}
Then we find that
\[
\tilde{g}^{ij}\frac{C_{j}}{C_{0}}\equiv\frac{C^{0}C^{i}}{C^{\lambda}%
C_{\lambda}}%
\]
The principal terms 4.12 are therefore
\begin{equation}
\frac{\mu_{p}^{\prime}\tilde{\partial}_{0}U_{0}}{C^{\lambda}C_{\lambda}%
}+\tilde{g}^{ij}\tilde{\partial}_{i}U_{j}%
\end{equation}

The matrix of the coefficients of the derivatives $\tilde{\partial}_{\alpha} $
in the equations deduced from the system 4.2, 4.7 is

\begin{center}
$\left(
\begin{array}
[c]{cccc}%
-\frac{\mu_{p}^{\prime}}{C^{\lambda}C_{\lambda}}\tilde{\partial}_{0} &
-\tilde{\partial}^{1} & -\tilde{\partial}^{2} & -\tilde{\partial}^{3}\\
-\tilde{\partial}_{1} & \tilde{\partial}_{0} & 0 & 0\\
-\tilde{\partial}_{2} & 0 & \tilde{\partial}_{0} & 0\\
-\tilde{\partial}_{3} & 0 & 0 & \tilde{\partial}_{0}%
\end{array}
\right)  $
\end{center}

it is symmetrized by taking the product with the $4\times4$ matrix

\begin{center}
$\left(
\begin{array}
[c]{cc}%
1 & 0\\
0 & \tilde{g}_{ij}%
\end{array}
\right)  .$
\end{center}

\subsection{Hyperbolicity.}

A FOS system is hyperbolic with respect to the space slices $x^{0}=$constant
if the system is still symmetric when written with the usual partial
derivatives, and is such that the corresponding matrix $M^{t}$ of the
coefficients of derivatives $\frac{\partial}{\partial t}$ is positive
definite. It admits then an energy inequality for a positive definite energy
relative to the space slices. This fact is of physical interest and also for
the proof of existence of solutions of the Cauchy problem, at least of
solutions local in time,

In the case we are considering the matrix $\tilde{M}^{0}$ is diagonal, with
positive elements if $\mu_{p}^{\prime}>0$ and $C^{\alpha}$ is timelike. The
principal matrix written with natural coordinates is also symmetrizable. The
corresponding matrix $M^{t}$ is not diagonal, and it is not obvious that it is
positive definite. In fact it will be so only if $\mu_{p}^{\prime}\geq1$. We
will prove the following lemma.

\begin{lemma}
The system 4.2, 4.7 is FOSH if $\mu_{p}^{\prime}\geq1$ and $C$ is timelike.
\end{lemma}

\begin{proof}
It is simpler to compute directly the energy inequality for the considered
system: its positivity is equivalent to the positivity of the matrix $M^{t}$.
Multiplying 4.2. and 4.7 respectively by $U_{0}$ and $\tilde{g}^{ij}U_{j}$
gives equations of the form
\begin{equation}
\frac{1}{2}\frac{\mu_{p}^{\prime}\tilde{\partial}_{0}(U_{0})^{2}}{f^{2}%
}-\tilde{g}^{ij}U_{0}\tilde{\partial}_{i}U_{j}=U_{0}F_{0}%
\end{equation}%
\[
\tilde{g}^{ij}U_{j}\tilde{\partial}_{0}U_{i}-\tilde{g}^{ij}U_{j}%
\tilde{\partial}_{i}U_{0}=\tilde{g}^{ij}U_{j}F_{i}%
\]
where the $F_{\alpha}$ contain only non differentiated terms. We add these two
equations, replace the operators $\tilde{\partial}$ by the operators
$\partial$ and carry out some manipulations using the expression for
$\tilde{g}^{ij}$ and the Leibniz rule. We obtain that
\[
\partial_{0}\mathcal{F}+\bar{\nabla}_{i}\mathcal{H}^{i}=Q_{F}%
\]
The function $Q_{F}$ is a quadratic form in $C_{\gamma\alpha}$ and the Riemann
tensor, while $\mathcal{H}^{i}$ and $\mathcal{F}$ are given by the following
expressions;
\[
\mathcal{H}^{i}=\frac{1}{C^{0}}\{\frac{1}{2}C^{i}(\frac{\mu_{p}^{\prime}%
}{f^{2}}U_{0}^{2}+\tilde{g}^{ij}U_{i}U_{j})-\tilde{g}^{ij}U_{0}U_{j}\},
\]%
\[
\mathcal{F}\equiv f^{-2}[(\mu_{p}^{\prime}-1)U_{0}^{2}+(U_{0}-C^{i}U_{i}%
)^{2}]+g^{ij}U_{i}U_{j}%
\]
The energy of the dynamical acceleration $\nabla C$ relative to the space
slices is the quadratic form $\mathcal{F}.$ It is positive definite if
$\mu_{p}^{\prime}\geq1$ and $C^{\alpha}$ is timelike.
\end{proof}

\begin{remark}
The system is hyperbolic in the sense of Leray if $\mu_{p}^{\prime}>0,$ but
the submanifolds $x^{0}=$constant are 'spacelike' with respect to the fluid
wave cone only if the fluid sound speed is less than the speed of light, i.e.,
$\mu_{p}^{\prime}\geq1.$
\end{remark}

\section{Bel-Robinson type energy of the system.}

The Bianchi system 3.1,3.2 and analogous equations obtained by replacing
$(0j)$ by $(ij)$, together with the system 4.2, 4.7. satisfied by the
dynamical acceleration constitute a FOSH system if $\mu_{p}^{\prime}\geq1. $
Its Bel - Robinson type energy (superenergy) density on a space slice is the
sum of the Bel - Robinson energy density of the gravitational field, and the
energy density of the dynamical acceleration:
\[
\mathcal{E\equiv B+F}.
\]
Using the expression of $\partial_{0}$ and the mean extrinsic curvature
$\tau\equiv g^{ij}K_{ij}$ of the space slices $S_{t},$ whose volume element we
denote by $\mu_{\bar{g}_{t}},$ we obtain an integral equality whose right hand
side couples gravitational and fluid superenergies:
\[
\int_{S_{t}}\mathcal{E\mu}_{\bar{g}}=\int_{S_{t}}\mathcal{E\mu}_{\bar{g}}%
+\int_{t_{0}}^{t}\int_{S_{\theta}}\{-N\tau\mathcal{E}+Q_{G}+\mathcal{S}%
+Q_{F}\}\mu_{\tilde{g}}d\theta
\]
The scalars $Q_{G},\mathcal{S}$ and $Q_{F}$ can be estimated in terms of
$\mathcal{E}$ and thus lead to a linear inequality for $\mathcal{E}$,
permitting the estimate of its growth with time.

\textbf{Aknowledgements. }We thank Cornell University and the NSF of the USA
for making possible a visit to Cornell by YCB. JWY also thanks the NSF (grant
n$%
%TCIMACRO{\UNICODE{0xb0}}%
%BeginExpansion
{{}^\circ}%
%EndExpansion
$ PHY- 9972582) for continuing support.

\textbf{References.}

Leray J. 1953 Hyperbolic differential equations I.A.S, Princeton

Foures (Choquet) - Bruhat Y. 1958 Bull Soc Math France 86, p. 155-175.

Bel L.1958, C.R. Acad. Sciences Paris.246 , 3105-3107.

Lichnerowicz A. 1964, Annales IHES n$%
%TCIMACRO{\UNICODE{0xb0}}%
%BeginExpansion
{{}^\circ}%
%EndExpansion
10.$

Choquet-Bruhat Y 1966, Comm Math Phys 3, 334-357

Friedrichs K.O. 1969, Fluid and magnetofluids, Colloque CNRS 1969

Ruggeri T, Strumia A, Ann. 1981, Ann. H. Poincare 34, 65-84

Anile M. 1982 Relativistic fluids and magneto fluids Camb. Univ. press.

Rendall A. 1992 J. Math Phys 33, 1047-1053.

Friedrich H. 1996 Class. Quantum Grav 13, 1451-1459.

Choquet-Bruhat Y, York J, 1997, Banach cent. pub. 41-1\ 119-131.

Friedrich H. 1998, Phys Rev D 57, 2317-2322.

Choquet - Bruhat Y. and York J. W. Top Met. in Non lin. An. 2001.

\bigskip

YCB. LPTL, Universit\'{e} Paris 6, 4, 75252, Paris. YCB@ccr.jussieu.fr

JWY. Physics Department, Cornell University, Ithaca, NY, 14853-6801, USA. York@astro.cornell.edu
\end{document}